# Data Descriptor Template

> **Scope Guidelines**
>
> **Data Descriptors.** This dataset provides a specialized resource for artificial intelligence applications in traditional Chinese medicine (TCM) research, specifically focusing on diagnostic image analysis. The collection consists of 6,719 clinically annotated tongue images representing 20 distinct pathological categories, with all labels validated by experienced TCM practitioners.
>
> Primarily designed for researchers developing deep learning models for automated tongue diagnosis, this dataset also serves as a robust benchmark for evaluating computer vision algorithms in complex medical imaging tasks. The unique characteristics of tongue diagnosis—including subtle color/texture variations—present challenging test cases for assessing model generalizability in real-world clinical scenarios.

## Title

*TCM-Tongue: A Standardized Tongue Image Dataset with Pathological Annotations for AI-Assisted TCM Diagnosis*

## Authors

Xuebo Jin[1], Longfei Gao[1], Anshuo Tong[1], Zhengyang Chen[1], Jianlei Kong[1], Ning Sun[2], Huijun Ma[1], Qiang Wang[3], Yuting Bai[1], Tingli Su[1]

## Affiliations

1. School of Computer and Artificial Intelligence, Beijing Technology and Business University, Beijing 100048, China.
2. Xiyuan Hospital of China Academy of Chinese Medical Sciences, No.1, R. Xiyuangcaochang, District Haidian, Beijing, 100091, China.
3. Peking University People's Hospital, Beijing, 100044, China.

corresponding author(s):
Jianlei Kong (kongjianlei@btbu.edu.cn)
Ning Sun(sunning_729@163.com)

## Abstract

Traditional Chinese medicine (TCM) tongue diagnosis, while clinically valuable, faces standardization challenges due to subjective interpretation and inconsistent imaging protocols, compounded by the lack of large-scale, annotated datasets for AI development. To address this gap, we present the first specialized dataset for AI-driven TCM tongue diagnosis, comprising 6,719 high-quality images captured under standardized conditions and annotated with 20 pathological symptom categories (averaging 2.54 clinically validated labels per image, all verified by licensed TCM practitioners). The dataset supports multiple annotation formats (COCO, TXT, XML) for broad usability and has been benchmarked using nine deep learning models (YOLOv5/v7/v8 variants, SSD, and MobileNetV2) to demonstrate its utility for AI development. This resource provides a critical foundation for advancing reliable computational tools in TCM, bridging the data shortage that has hindered progress in the field, and facilitating the integration of AI into both research and clinical practice through standardized, high-quality diagnostic data.

# Background & Summary

### Introduction.

Traditional Chinese Medicine (TCM), with its millennia-old history, is rooted in a holistic philosophy that emphasizes the harmony between nature and the human body, integrating ecological, psychological, cultural, and scientific perspectives into a unified system of diagnosis and treatment. Unlike modern Western medicine, TCM relies heavily on observational techniques, among which tongue diagnosis is a cornerstone of the "Four Diagnostic Methods" [1]. This process traditionally depends on a practitioner's subjective interpretation of visual features such as color, texture, and coating—a method inherently limited by human perception and variability.

Recent advances in artificial intelligence (AI), particularly deep learning-based image analysis, offer unprecedented opportunities to modernize TCM diagnostics. By automating tongue image assessment, AI can enhance objectivity, consistency, and scalability in TCM practice.

### Challenges.

However, the integration of AI into tongue diagnosis faces three critical challenges:

1. **Data Scarcity & Fragmentation:** Unlike standardized medical imaging datasets (e.g., X-rays or MRIs) collected by standard instruments, tongue diagnosis data remains siloed within individual practitioners' records, limiting public availability. Most existing datasets are small, non-standardized, or lack annotations suitable for AI training.
2. **Acquisition Inconsistencies:** Even when data is collected, variations in lighting, camera angles, and patient conditions introduce noise, reducing algorithmic robustness[2]. Without standardized protocols, aggregated datasets fail to meet the uniformity required for reliable deep learning applications.
3. **Labelling Complexity:** TCM diagnostics focus on "symptom patterns" rather than disease entities, requiring annotations that reflect TCM theory (e.g., "pale tongue with white coating" indicating Qi deficiency). Such labels demand expert knowledge and must align with both TCM principles and modern object-detection frameworks (e.g., bounding boxes for tongue regions or sub-features).

To date, no large-scale, publicly available tongue image dataset meets these requirements. Existing resources either lack TCM-compliant annotations or sufficient sample diversity for deep learning. Addressing this gap necessitates a rigorously collected dataset with: (1) standardized imaging protocols, (2) clinically validated labels from experienced TCM practitioners, and (3) compatibility with mainstream AI tools. This paper introduces the first such dataset, designed to bridge TCM diagnostics and AI research while preserving the theoretical integrity of traditional medicine.

### Related Image Datasets Summary.

The proliferation of image detection tasks across diverse application domains has necessitated the development of comprehensive and representative image datasets. These datasets serve as benchmarks for evaluating and enhancing the performance of image detection algorithms.

The UNISA2020 dataset leads source camera identification with real images from identical digital cameras, essential for evaluating Source Camera Identification algorithms (Andrea Bruno et al[3].). In medical imaging, particularly laparoscopic hysterectomy, AutoLaparo addresses the need for high-quality, multi-task labeled data in computer-assisted surgery (Ziyi Wang et al[4].). The BAID dataset of 3000 backlit images aids in learning resilient enhancers for backlit photos (Xiaoqian Lv et al[5].). The Whole Abdominal Organ Dataset (WORD) meets the demand for medical segmentation data, with 150 CT volumes annotating 16 abdominal organs, aiding research and clinical work in organ segmentation (Xiangde Luo et al[6].). For floorplan analysis, the CubiCasa5K dataset enhances capabilities with 5000 samples annotated into 80+ categories (Ahti Kalervo et al[7].). In ultrasound video analysis for breast lesion detection, a dataset of 188 annotated videos supports training advanced models like CVA-Net (Zhi Lin et al[8].). The SegPC-2021 challenge released 775 images of Multiple Myeloma plasma cells, advancing digital pathology tools and cell segmentation (Anubha Gupta et al[9].).

These datasets represent significant progress across various facets of image detection, each addressing unique challenges within their respective fields. Developing these datasets enhances both research and practical application in image detection by advancing the capabilities of machine learning models by providing comprehensive and diverse training datasets. Although these medical imaging datasets have contributed to the discipline, they often exhibit a concentrated utility, primarily serving specific medical specializations. While this specificity is advantageous for specialized research, it may limit the applicability of the models developed for broader medical or traditional tactile fields. Furthermore, the inherent challenges associated with labeling, including the requirement for specialized medical expertise, can hinder the expansion of the dataset and the implementation of more generalized image detection models.

To the best of our knowledge, there is currently no formally published and publicly available AI-ready dataset specifically designed for TCM tongue diagnosis research. The lack of standardized, large-scale tongue image datasets with comprehensive annotations remains a significant gap in the field of AI-assisted TCM diagnostics. This limitation hinders the development and benchmarking of robust machine learning models for automated tongue analysis in clinical applications.

**Our Contribution.**

This study presents a novel, expert-annotated dataset designed to bridge the gap between TCM tongue diagnosis and modern deep learning methodologies. Addressing key challenges—such as diagnostic subjectivity, environmental variability, and data scarcity—we introduce a meticulously curated collection of 6,719 labeled tongue images, spanning 20 distinct symptom categories reflective of TCM theory. With an average of 2.54 diagnostic labels per image, this dataset captures the granularity and clinical nuance essential for accurate TCM assessment, offering a rare resource for AI-driven research in traditional medicine.

Unlike conventional medical imaging datasets, our work fills a critical void by digitizing a centuries-old diagnostic tradition. Each annotation was rigorously validated by authoritative TCM practitioners, ensuring adherence to classical symptom differentiation while meeting the technical demands of deep learning frameworks. By standardizing tongue image acquisition and labeling, we mitigate longstanding issues of inconsistency, enabling robust model training for tasks such as feature segmentation, symptom classification, and automated diagnosis.

Beyond facilitating AI applications in TCM, this dataset invites broader exploration of cross-cultural medical AI. Traditional diagnostics—often marginalized in digital health initiatives—present unique challenges for computer vision, including subtle feature variations (e.g., coating thickness, color gradations) and context-dependent interpretations. Here, we demonstrate how AI can adapt to these complexities, extending the versatility of image-based diagnostics beyond Western biomedicine.

The implications are twofold:

1. **Advancing TCM Digitization:** By providing a high-quality benchmark dataset, we accelerate the integration of AI into TCM practice, enhancing objectivity and scalability while preserving its holistic principles.
2. **Expanding AI's Medical Scope:** This work will extend the dominance of Western-centric medical AI, showcasing how deep learning can embrace diverse diagnostic traditions. Future applications may include personalized TCM diagnostics, telemedicine platforms, and hybrid AI-human decision systems.

In summary, our dataset not only pioneers the computational modernization of TCM but also underscores AI's potential to unify traditional and contemporary medical paradigms. By leveraging machine learning to decode ancient diagnostic wisdom, we open new pathways for interdisciplinary innovation in global healthcare.

# Methods

## Overview.

To maximize the research utility of this dataset in AI-augmented TCM tongue diagnosis, our study adopts a dual-pronged methodological approach:

1. **Standardized Image Acquisition Protocol:** We designed a dedicated hardware system for tongue image capture, ensuring consistent lighting conditions, angle alignment, and color calibration to minimize environmental variability. Further, hospital practitioners underwent rigorous training protocols to standardize data collection procedures, guaranteeing uniformity across all samples. This mitigates inter-operator discrepancies and enhances dataset reliability for downstream AI applications.
2. **Expert-Annotated, AI-Ready Labelling Framework:** To ensure both clinical authenticity and technical applicability, the diagnostic labels were meticulously curated under the guidance of renowned TCM physicians, achieving a dual alignment that captures essential TCM-specific nuances while maintaining deep learning compatibility. The annotations preserve classical TCM diagnostic markers—including tongue coating texture, color gradations, and fissure patterns—rooted in traditional theory, while simultaneously being structured as multi-label classifications and segmentation masks optimized for training modern neural networks such as YOLOs and vision transformers. This approach effectively bridges TCM's holistic diagnostic framework with the computational requirements of AI, enabling the development of models that are both clinically meaningful for traditional medicine and technically robust for machine learning applications.

By integrating domain expertise with technical standardization, our dataset not only supports advanced AI research but also upholds the integrity of TCM's diagnostic traditions.

## Methodological Implementation of the Tongue Diagnosis Capture System.

To ensure rigorous standardization in data acquisition, we developed a purpose-built tongue imaging system (Figure 1) based on the PyQT framework. The system integrates a synchronized dual-camera array with precision-calibrated illumination modules to achieve consistent imaging conditions. The primary wide-angle imaging unit incorporates intelligent facial proximity detection within a 30-50cm operational range, while its embedded ResNet-50 convolutional neural network performs real-time demographic analysis through advanced facial recognition. This initial profiling ensures subject eligibility and triggers subsequent imaging protocols.

Upon successful facial verification, the system activates its secondary telephoto imaging component through an event-triggered mechanism. This high-resolution module employs 5x optical zoom with focus stacking capability, enabling sub-100μm resolution for detailed tongue surface topography reconstruction. To optimize subject positioning, an adaptive voice guidance protocol provides six-degree-of-freedom adjustment instructions, while real-time quality assessment monitors focus peaking, tongue coverage, and motion artifacts. Cross-polarized, spectrally calibrated LED arrays (D65 standard illuminant) automatically adjust intensity between 500-1500 lux to minimize specular reflections and maintain consistent lighting across captures.

The complete acquisition cycle is completed within 3-8 seconds, with all imaging parameters automatically logged in DICOM-compatible metadata. This integrated approach not only ensures clinical-grade image quality but also guarantees research-grade reproducibility, making the dataset robust for downstream AI analysis. By combining automated demographic screening, adaptive positioning guidance, and high-precision optical capture, the system bridges clinical workflow efficiency with the technical demands of machine learning applications. Figure 2 shows examples of images captured by the tongue diagnosis capture system.

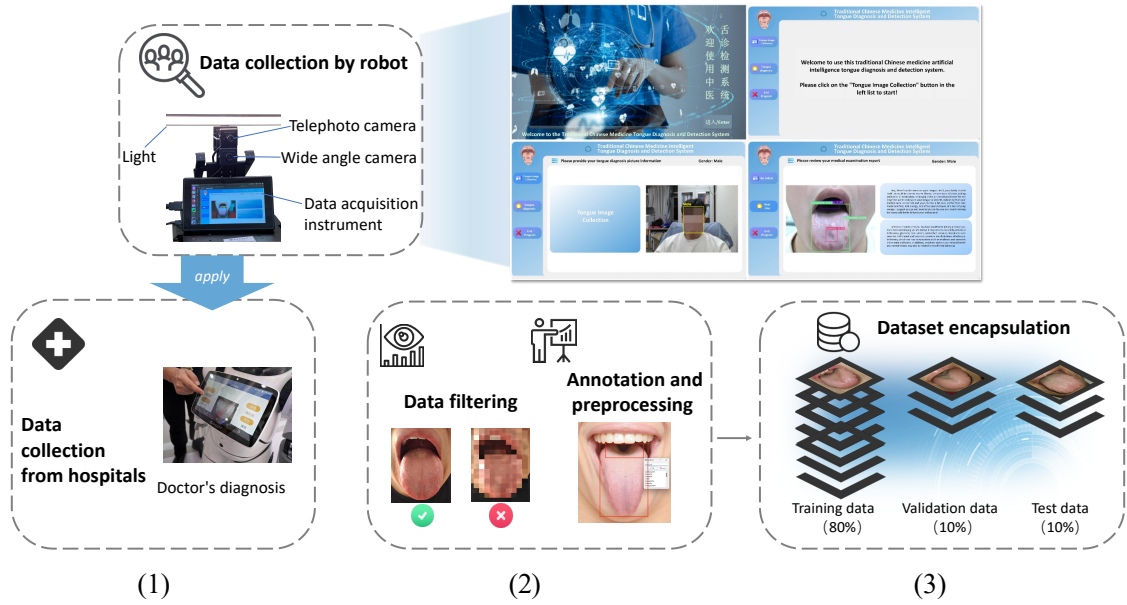

**Fig.1 Standardized workflow for tongue image acquisition and dataset construction.** The protocol initiates with participant registration and informed consent procedures, followed by automated positioning guidance using our specialized imaging device. The integrated system concurrently performs three key functions: (1) demographic profiling via facial recognition technology, (2) standardized image acquisition under controlled lighting conditions (D65 standard illuminant), and (3) systematic archiving of acquired tongue images. The complete dataset was partitioned into training (80%), validation (10%), and test (10%) subsets using stratified random sampling to maintain diagnostic category distributions.

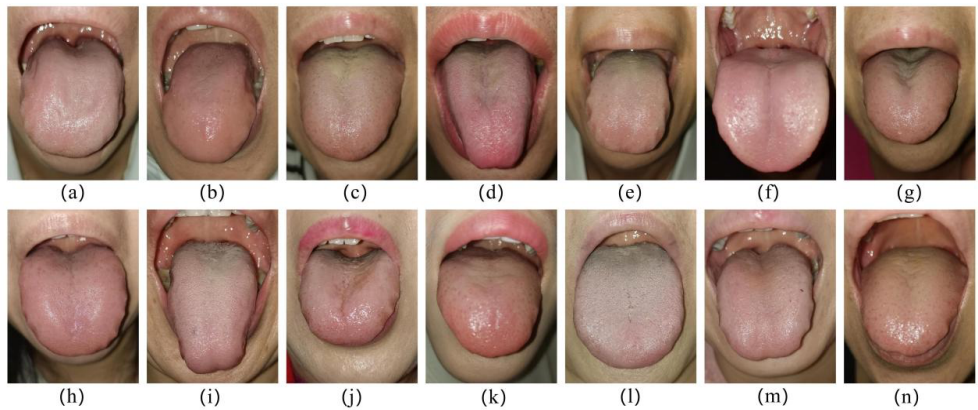

**Fig.2 Images captured by the tongue diagnosis capture system.**

## Label Selection and Annotation.

The images acquired through the standardized tongue diagnosis capture system will subsequently undergo multi-label annotation by TCM practitioners. Our label selection approach is fundamentally rooted in three core principles of TCM: (1) the preventive orientation that focuses on detecting subclinical pathological patterns before they manifest as overt disease; (2) the holistic assessment framework that evaluates tongue features not as isolated markers but as interconnected systems reflecting the body's dynamic equilibrium; and (3) the semantic continuity that deliberately preserves TCM's original diagnostic taxonomy to maintain conceptual integrity with centuries of clinical observation and theoretical development. These principles work synergistically to ensure our computational models capture the essence of TCM diagnosis - where early intervention potential, systemic correlation interpretation, and terminological precision together form an inseparable foundation for accurate pattern differentiation and clinical decision-making.

Specifically, we selected 20 TCM tongue diagnosis feature annotations based on classical literature and clinical consensus from clinical practitioners[10][11]. This system maintains precise alignment with original TCM pathophysiological concepts, enabling accurate pattern differentiation in accordance with TCM's syndrome classification framework. By adhering to these principles, the annotation system ensures that computational models derived from it uphold the theoretical integrity of TCM while translating traditionally qualitative observations into quantifiable data. The label set encompasses all major diagnostic dimensions of tongue examination, including surface characteristics such as coating and moisture, morphological changes like teeth marks and cracks, color variations ranging from pale to crimson, and regional abnormalities corresponding to specific organ systems.

As listed in Table 1, the label systematically categorizes both global tongue characteristics (e.g., color, texture, and morphology) and localized pathological markers (e.g., organ-specific depressions or protrusions). The dataset includes fundamental diagnostic indicators such as healthy tongue (jiankangshe), color variations (e.g., hongshe, zishe), texture abnormalities (e.g., liewenshe, chihenshe), and region-specific features (e.g., shenquao, gandantu). Each label is numerically indexed (0-19) to facilitate machine learning applications in automated tongue image analysis. This structured dataset serves as a valuable resource for advancing research in digital TCM diagnostics and AI-assisted pattern recognition.

The annotation approach comprises three key elements: (1) label category classification, (2) center coordinates of the bounding box, and (3) dimensional parameters (width and height). A comprehensive quality control system spans the entire workflow from annotation to release. Initial annotations by trained technicians undergo review by TCM practitioners with 5+ years of experience, with critical cases requiring consensus from expert panels. Generated using the industry-standard LabelImg annotation tool, the dataset provides dual-format annotation files: PASCAL VOC XML files that preserve hierarchical metadata of diagnostic features, and YOLO TXT files with normalized coordinates for real-time detection requirements.

As illustrated in Fig. 3, the labeling system employs a dual-category approach: Global Labels, which capture the tongue's holistic characteristics, and Local Labels, which focus on specific regional features. This hierarchical annotation framework enables comprehensive analysis by simultaneously representing macroscopic conditions and microscopic pathological details.

Fig. 4 visually demonstrates the annotation scheme, distinguishing between global and local labels through a standardized color-coding system. Global labels, marked by a red box for white-coated tongue (baitaishe) and an orange-yellow box for red tongue (hongshe), capture the tongue's overall appearance, with both labels occupying the same spatial region to reflect coexisting conditions. Local labels, highlighted in blue for cracked tongue (liewenshe) and green for dentate tongue (chihenshe), pinpoint specific pathological features, enabling fine-grained analysis. This dual-level annotation approach ensures comprehensive representation of both macroscopic and microscopic diagnostic indicators while maintaining visual clarity for automated detection.

Table 1 Twenty Clinically Validated Tongue Feature Categories for TCM Pattern Differentiation

| Label Name | Category | Label Type | Detailed Explanation |
| --- | --- | --- | --- |
| jiankangshe | Healthy Tongue | Global Label | Presents as light red with moist texture, moderate size, and thin white coating. Characterized by absence of stickiness/greasiness or morphological abnormalities. Represents balanced qi-blood status in healthy individuals. |
| botaishe | Tongue with Peeling Coating | Local Label | Exhibits partial/complete coating loss ("map tongue"), typically white-greasy. Pathognomonic for spleen-stomach deficiency (qi decline) and yin-fluid depletion. |
| hongshe | Red Tongue | Global Label | Erythema indicates heat syndrome progression (excess fire/qi). Hue intensity correlates with |

| | | | pathological heat severity (e.g., crimson suggests extreme heat). |
|---|---|---|---|
| zishe | Purple Tongue | Global Label | Reflects cold/heat syndromes or blood stasis. Modern pathophysiology involves microcirculatory disturbances, hypoxia, or metabolic disorders. |
| pangdashe | Chubby Tongue | Global Label | Macroglossia with potential protrusion. Signifies cold-damp retention from spleen/kidney yang deficiency (water-dampness internal accumulation). |
| shoushe | Thin Tongue | Global Label | Atrophic morphology with reduced volume. Indicates blood/yin deficiency syndromes (e.g., spleen deficiency, essence depletion). |
| hongdianshe | Red Dot Tongue | Local Label | Fungiform papillary hyperemia presenting as punctate lesions. Associated with heat-toxin accumulation or blood stasis. |
| liewenshe | Cracked Tongue | Local Label | Variably patterned fissures/grooves. Signifies essence-blood depletion or chronic yin deficiency. |
| chihenshe | Dentate Tongue | Local Label | Lingual scalloping from dental pressure. Pathognomonic for spleen qi/yang deficiency with dampness retention. |
| baitaishe | White Coating Tongue | Global Label | Physiological: thin-white. Pathological: thick-white suggests cold-damp obstruction or food stagnation. |
| huangtaishe | Yellow Coating Tongue | Global Label | Transition from white indicates heat progression (e.g., stomach heat, inflammatory processes). |
| heitaishe | Black Coating Tongue | Global Label | Critical sign of extreme heat/cold. Requires urgent intervention (e.g., heat exhaustion or yang collapse). |
| huataishe | Smooth Coating Tongue | Local Label | Hyperhydrated surface with drooling tendency. Indicates yang deficiency failing to transform dampness (common in spleen-kidney yang deficiency). |
| shenquao | Tongue with Sunken Kidney Area | Local Label | Depression at tongue root. Reflects kidney qi deficiency with associated lumbago/asthenia. May correlate with adrenal fatigue in integrative medicine. |
| shenqutu | Tongue with Protruding Kidney | Local Label | Bulging lesions suggest kidney heat/toxin accumulation. Differential includes chronic nephropathies or metabolic disorders. |
| gandanao | Tongue with Concave Liver and Gallbladder Area | Local Label | Lateral tongue depressions. Signifies liver-blood deficiency or qi stagnation (e.g., chronic stress or sleep deprivation effects). |
| gandantu | Tongue with Protruding Liver and Gallbladder Area | Local Label | Nodular elevations indicate liver qi stasis or biliary disorders (e.g., gallstones, fatty liver disease). |
| piweiao | Tongue with Sunken | Local Label | Mid-tongue concavity. Pathognomonic for spleen-stomach qi/yin deficiency with possible malabsorption. |

| | Spleen and Stomach Area | | |
|---|---|---|---|
| xinfeiao | Tongue with Sunken Heart and Lung Area | Local Label | Anterior depression suggests cardiopulmonary qi deficiency or chronic hypoxia (e.g., sleep apnea effects). |
| xinfeitu | Tongue with Protruding Heart and Lung Area | Local Label | Tip elevation reflects lung yin deficiency or heart fire. Associated with chronic respiratory/circulatory stress. |

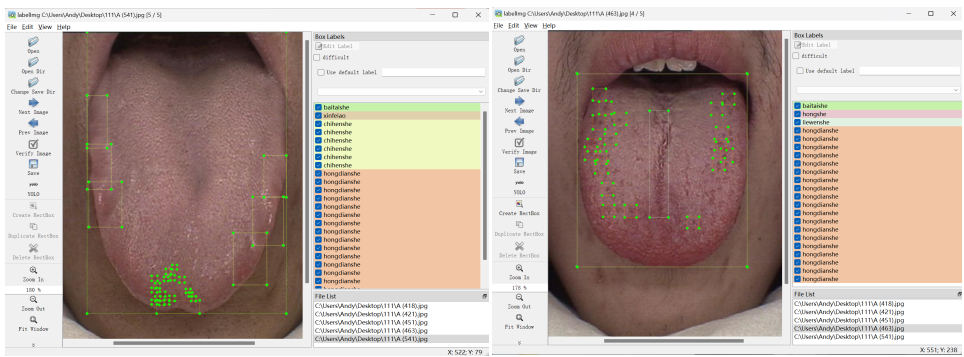

Fig.3 Workflow of multi-level annotation for computerized tongue image analysis

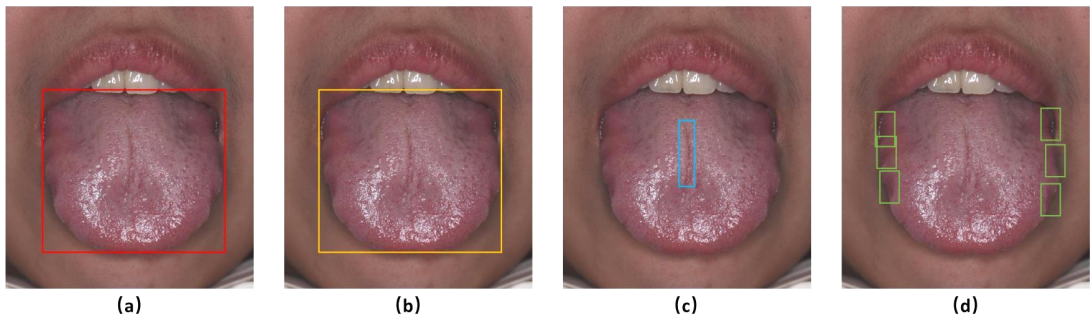

Fig.4 Color-coded annotation scheme for tongue features: (a) red box - white-coated tongue(global labels), (b) orange-yellow box - red tongue (global labels, same position as white-coated label); (c) blue box - cracked tongue(local labels), (d) green box - dentate tongue(local labels).

## Data Records

Figure 5 shows the situation of the labels on the image of the tongue. We find that, unlike other image target detection, the labels for the target not only encompass the overall condition of the entire surface of the tongue but also emphasize the specific local areas, reflecting the observational characteristics of the tongue in TCM. Fig. 6 illustrates the distribution of image quantities for each label, showing an average of 2.54 labels per tongue image.

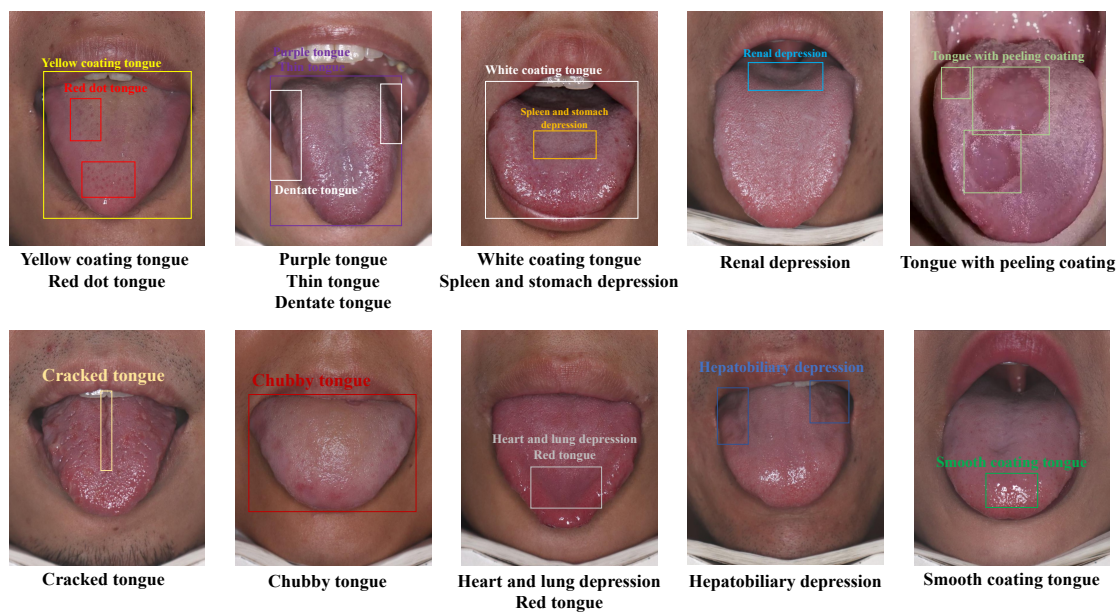

Fig.5  Typical examples of image dataset for tongue diagnosis detection

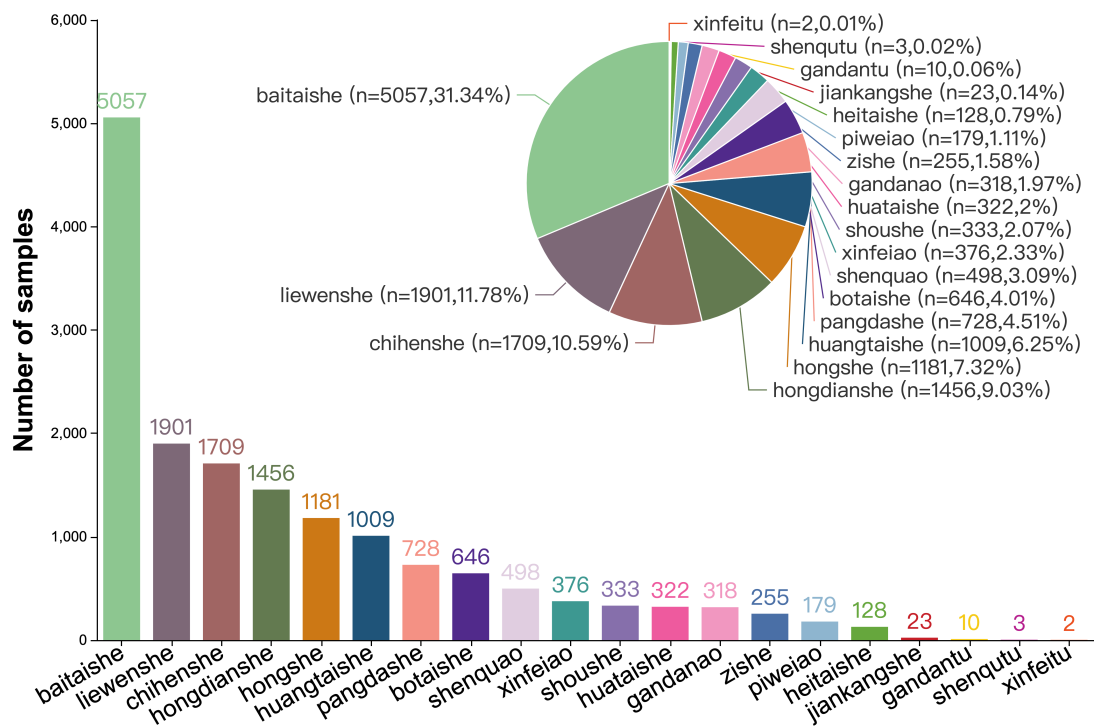

Fig.6  Distribution of labels in the tongue diagnosis dataset

Comprising 6,719 clinically validated tongue images, the dataset follows machine learning best practices in its division: 5,594 images (82.3%) for training, 572 (8.4%) for validation, and 553 (8.1%) for final evaluation. A specialized subset containing 10% "challenging cases" represents borderline tongue manifestations in TCM differential diagnosis, providing valuable test scenarios for model robustness.

The complete dataset is available through Baidu Cloud, please click on the GitHub website to download(https://github.com/btbuIntelliSense/Intelligent-tongue-diagnosis-detection-dataset),

including high-resolution source images, multi-format annotation files, detailed technical documentation, and preprocessing scripts. To facilitate replication studies, sample code for PyTorch and TensorFlow implementations is provided, covering complete workflows from data loading and augmentation to evaluation.

## Technical Validation

To comprehensively evaluate the dataset, all the code used for training and testing models in this dataset was run in a Linux environment, Ubuntu 20.04, CUDA version 11.4, and the experimental framework used was PyTorch. The processor model is Intel Core i7-6800K 3.40GHz, and the graphics card model is three Nvidia GeForce GTX 1080 Ti, with 11GB of graphics memory and 32GB of memory.

The selected algorithm is also a classic detection algorithm in both one-stage and two-stage detection networks. The precision, recall, and mAP are used as evaluation indicators [12], and their calculation formulas are shown in Equations (1)-(5). In addition, mAP0.5 represents the mAP value at an intersection over union(IoU[13]) of 0.5, and mAP0.5:0.95 represents the average value of each mAP with an increase of 0.05 in the intersection over union(IoU[13]) between 0.5 and 0.95. In the formula, TP[12] represents the number of true positive samples, FP represents the number of false positive samples, FN[12] represents the number of false negative samples, and C represents the number of categories:

$$\Pr ecision = \frac{TP}{TP+FP} \quad (1)$$

$$\mathrm{Re} call = \frac{TP}{TP+FN} \quad (2)$$

$$AP = \int_0^1 \Pr ecision\, d(\mathrm{Re} call) \quad (3)$$

$$mAP = \frac{\sum_{i=1}^c AP_i}{C} \quad (4)$$

$$IoU = \frac{TP}{TP+FP+FN} \quad (5)$$

Table 2 Performance Comparison of Deep Learning-Based Detectors on the Tongue Diagnosis Dataset

| Algorithm | Param /Million | Weight_Size /MB | P/% | R/% | mAP$_{0.5}$/% | mAP$_{0.50.95}$/% |
|---|---|---|---|---|---|---|
| YOLOv5s[14][15] | 7.0 | 14.5 | 39.61 | 41.26 | 31.83 | 26.31 |
| YOLOv5m[16] | 20.9 | 42.3 | 42.47 | 38.54 | 32.48 | 26.65 |
| YOLOv5l[17] | 46.2 | 92.9 | 43.76 | 40.29 | 34.57 | 27.33 |
| YOLOv7-tiny[18] | 6.03 | 12.3 | 40.77 | 46.18 | 31.93 | 25.09 |
| YOLOv7[19][20] | 37.2 | 74.9 | 47.23 | 37.12 | 34.82 | 28.38 |
| YOLOv8s[21][22] | 11.1 | 22.5 | 40.75 | 35.52 | 33.54 | 27.98 |
| YOLOv8m[23] | 25.9 | 52.1 | 40.36 | 37.84 | 34.77 | 28.20 |
| YOLOv8l[24] | 43.6 | 87.7 | 38.19 | 41.07 | 34.95 | 28.59 |
| SSD[25][26] | 34 | 102.7 | 38.72 | 34.75 | 28.86 | 22.90 |
| MobileNetV2[27] | 3.4 | 24.5 | 34.07 | 11.75 | 23.20 | 13.90 |

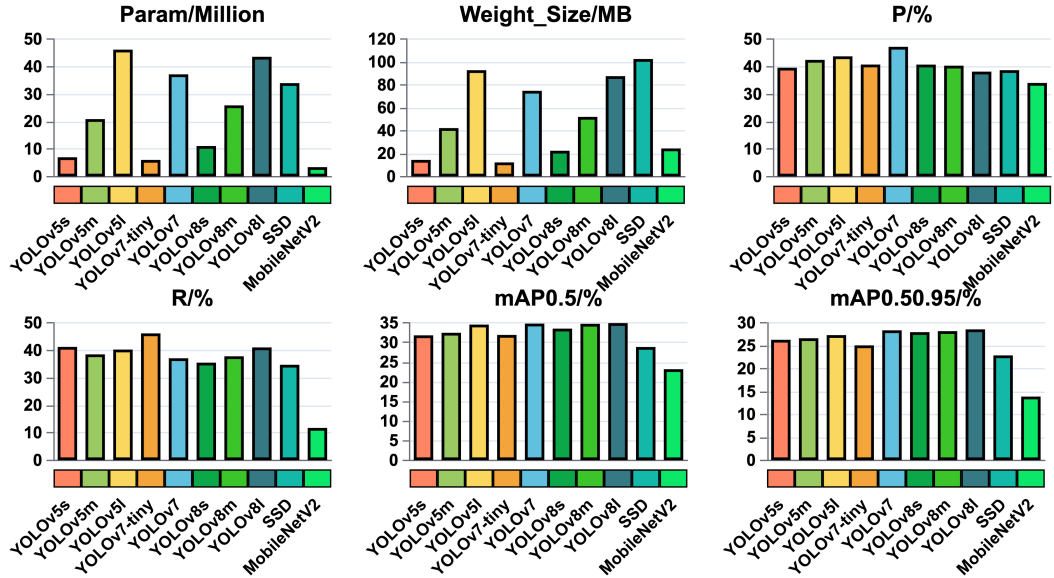

Fig.7  Bar Chart Visualization: Performance of Tongue Diagnosis Detectors

The study evaluates several deep learning-based object detection models, including various versions of YOLOv5, YOLOv7, YOLOv8, SSD, and MobileNetV2, on a tongue diagnosis dataset to determine the most suitable architecture for medical image analysis. Table 2 summarizes the performance comparison among deep learning-based detectors on the tongue diagnosis dataset, while Figure 7 provides a bar chart visualization of these results. Contrary to expectations, increasing model depth within the same family (e.g., from YOLOv5s to YOLOv5l) does not consistently improve accuracy or recall, with performance instead showing fluctuations or even degradation. This phenomenon likely stems from the tongue dataset being significantly smaller than standard benchmarks, such as COCO, causing deeper models with more parameters to overfit. Their redundant parameters also hinder generalization.

Among lightweight models, YOLOv8s emerges as particularly effective, achieving superior mAP0.5 (33.54%) and mAP0.5-0.95 (27.98%) compared to similar-sized models, such as YOLOv7-tiny and YOLOv5s, while maintaining a low parameter count of just 11.1 million. For applications requiring higher accuracy, YOLOv7 and the medium- and large-scale YOLOv8 variants deliver the best performance, with YOLOv8l achieving 34.95% mAP@0.5. Notably, YOLOv7 achieves competitive accuracy (34.82% mAP@0.5) with fewer parameters (37.2M) than YOLOv8l (43.6M), thanks to its efficient E-ELAN module, which optimizes gradient flow for improved feature learning. While YOLOv8 shows slight mAP improvements through its innovative C2F structure and Anchor-Free detection strategy, the practical advantages of YOLOv7's parameter efficiency and faster inference make it particularly suitable for tongue diagnosis applications. The poorer performance of SSD (28.86% mAP0.5) and MobileNetV2 (23.20% mAP0.5) highlights the limitations of older architectures and extremely lightweight designs for this complex medical imaging task. These findings suggest that for tongue diagnosis and similar specialized medical imaging tasks with limited datasets, mid-sized models like YOLOv7 or YOLOv8m offer the optimal balance between accuracy and computational efficiency, avoiding both the underperformance of overly simplistic models and the diminishing returns of excessively large architectures.

It is evident from Figures 8 and 9 that the deeper the network depth, the higher the mAP0.5 metric. In the first 100 rounds of training, the convergence speed of YOLOv7 tiny will be slower than YOLOv5s and YOLOv8s. YOLOv8m is even comparable to the average accuracy of YOLOv7 and YOLOv8l detectors, with larger parameter quantities than it. Due to its outdated network structure, SSD is even difficult to outperform detectors with much smaller parameter quantities. However, MobileNetV2 did not perform well in the first 100 epochs of training due to insufficient network depth.

As we know, mAP50-95 is an indicator obtained by averaging 10 mAP values with an interval of 5%, ranging from a mAP threshold of 50% to a mAP threshold of 95%. Compared to mAP50, it is more stringent and requires higher network detection capabilities. As shown in Figure 10, we compared the performance of non-lightweight networks on the tongue diagnosis dataset, with an evaluation metric of mAP50-95. It can be seen that the performance of YOLOv8m and YOLOv8l is very impressive, followed by YOLOv7. Considering the application methods of intelligent tongue diagnosis and the particularity of tongue diagnosis datasets, a balance between computing power and accuracy should be considered, because in most scenarios, the computing power of intelligent tongue diagnosis robots is not sufficient to support fast inference and testing, and relatively lightweight networks are more feasible. Further, it is necessary to ensure the accuracy of the detector to support the scientific rigor of tongue diagnosis results.

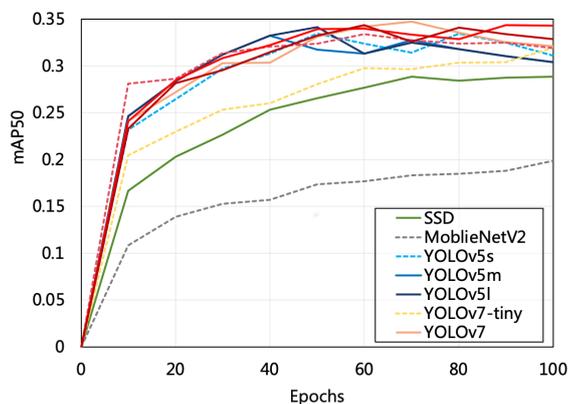

Fig.8  Epoch-wise mAP50 Evolution: Performance Benchmarking in Initial 100 Epochs

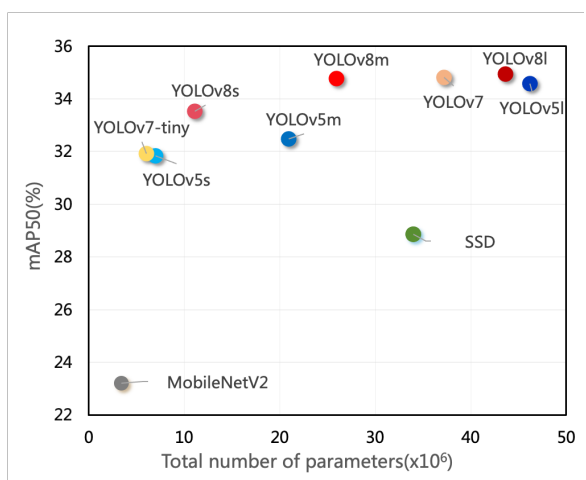

Fig.9  Joint Visualization of mAP50 Accuracy and Parameter Scale for Tongue Image Analysis

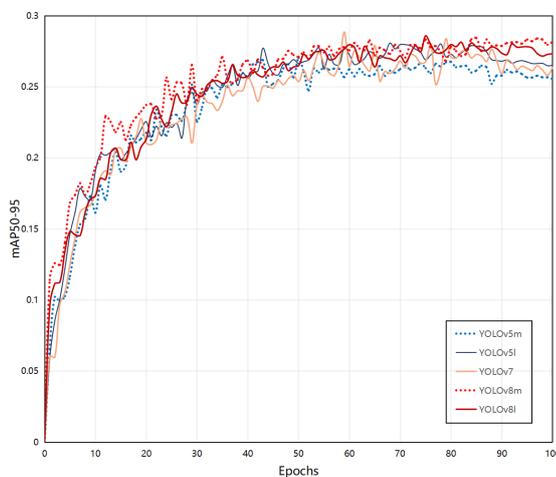

Fig.10  Comparison of mAP50-95 between medium and large detectors on the tongue diagnosis dataset

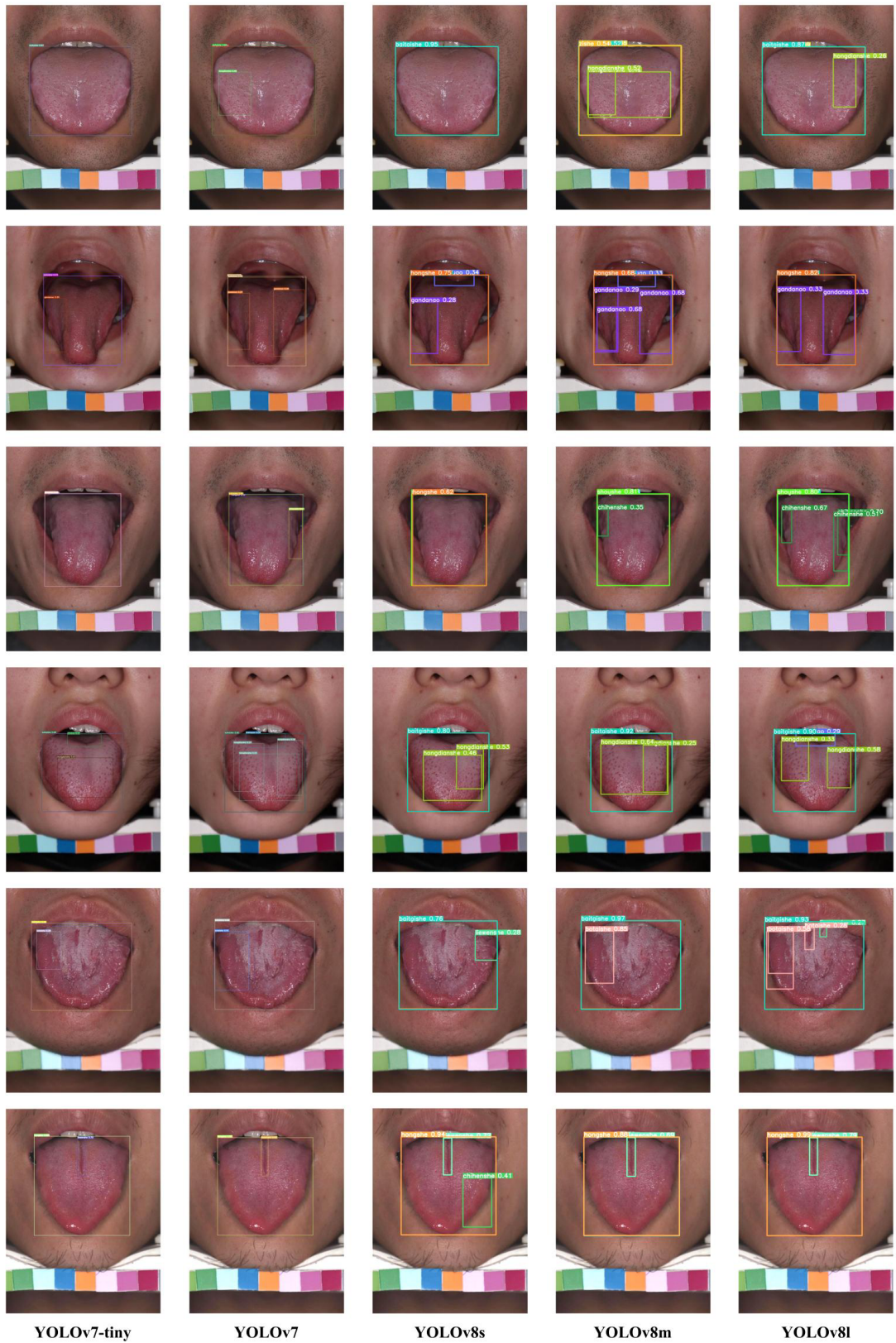

Fig.11 Benchmarking Object Detection Performance on Tongue Diagnosis Dataset

In Figure 11, we compared the results of the latest detection network trained on the tongue diagnosis dataset and tested on test cases. It is not difficult to see that YOLOv8l, as the model with the largest number of parameters, has more accurate and comprehensive detection and classification results. However, YOLOv7-tiny and YOLOv8s with smaller parameter quantities can reduce detection time and GPU memory consumption, but there are often false positives and missed detections, which are often caused by incomplete feature extraction and insufficient network fitting. Meanwhile, in the face of similar color features such as purple and red tongues, the classification performance of the above detectors still needs to be improved; Improving the detection performance of detectors is also crucial for irregular features such as cracked tongues and dentate tongues.

## Usage Notes

This dataset supports both object detection and classification tasks. In addition to the aforementioned YOLO series [28], compatible networks also include SD[29], Faster R-CNN[30], and Mask R-CNN[31] models. The data comes pre-processed and can be directly loaded using Python (OpenCV/PIL) or deep learning frameworks (PyTorch/TensorFlow).

Refer to the "Label Selection and Annotation" sections for detailed labeling criteria. The dataset is ready for immediate use - simply download and start training without additional preprocessing.

## Code Availability

Our work is currently open source and will continue to be revised and updated in the future. Currently, it has been updated to version 3.0, including the coco annotation format, the xml annotation format, and the YOLO annotation format. txt. These annotation data are all provided for free (https://github.com/btbuIntelliSense/Intelligent-tongue-diagnosis-detection-dataset/tree/main). Our dataset is provided through cloud storage, and there are also some simple instances in the demo, along with configuration files for the relevant models. We have also added a brief explanation for the dataset in README.md.

## Acknowledgements

This work was supported in part by the National Natural Science Foundation of China No. 62173007, 62203020, 62473008, 62433002, 62476014, Beijing Nova Program (20240484710), Project of Humanities and Social Sciences No. 22YJCZH006 (Ministry of Education in China, MOC), Beijing Scholars Program Grant 099, Project of ALL China Federation of Supply and Marketing Cooperatives (No. 202407), and Project of Beijing Municipal University Teacher Team Construction Support Plan (No. BPHR20220104).

## Author contributions

Xuebo Jin and Longfei Gao conceived this project. Anshuo Tong, Zhangyang Chen and Huijun Ma conducted data annotation, experimental calculations, and analysis. Qiang Wang and Tingli Su have provided technical and scientific input for the research on intelligent tongue diagnosis in traditional Chinese medicine. All authors provided feedback on the analysis. Yuting Bai and Jianlei Kong wrote the manuscript based on the opinions and revisions of all authors.

## Competing interests

The authors declare no competing interests.

## Additional information

Correspondence and requests for materials should be addressed to jinxuebo@btbu.edu.cn.